\documentclass[12pt]{article}
\usepackage{epsfig}
\usepackage{psfig}
\usepackage{amssymb}
 
\begin{document}

\begin{center}
\Large
{An ultra-relativistic outflow from a neutron star accreting gas from
  a companion}
\normalsize

-----------------------------------------------------------
\end{center}
\smallskip

\begin{center}
\Large
{Nature, 427, 222-224, 2004}
\normalsize

-----------------------------------------------------------
\end{center}
\smallskip

\noindent{\bf Rob Fender}, 
{\em Astronomical Institute `Anton Pannekoek', University of
Amsterdam, Kruislaan 403, 1098 SJ Amsterdam, The Netherlands}

\noindent{\bf Kinwah Wu}, 
{\em Mullard Space Science Laboratory, University College London,
Holmbury St Mary, Surrey, RH5 6NT, UK}

\noindent{\bf Helen Johnston}, 
{\em School of Physics, University of Sydney, NSW 2006, Australia}

\noindent{\bf Tasso Tzioumis}, 
{\em Australia Telescope National Facility, CSIRO, P.O. Box 76, Epping NSW 1710, Australia}

\noindent{\bf Peter Jonker}, 
{\em Institute of Astronomy, University of Cambridge, Madingley Road, Cambridge CB3 0HA, UK}

\noindent{\bf Ralph Spencer}, 
{\em University of Manchester, Nuffield Radio Astronomy Laboratories,
Jodrell Bank, Cheshire, SK11 9DL, UK}

\noindent{\bf Michiel van der Klis}, 
{\em Astronomical Institute `Anton Pannekoek', University of
Amsterdam, Kruislaan 403, 1098 SJ Amsterdam, The Netherlands}

\bigskip
\begin{center}
-----------------------------------------------------------
\end{center}
\normalsize

{\bf Collimated relativistic outflows, or jets, are amongst the most
energetic and relativistic phenomena in the Universe.  They are
associated with supermassive black holes in distant active galactic
nuclei (AGN) \cite{ost97}, accreting black holes and neutron stars in
binary systems \cite{mr99} and are believed to be responsible for
gamma-ray bursts \cite{sar99}. The physics of these jets,however,
remains something of a mystery in that their bulk velocities,
compositions and energetics remain poorly determined.  Here we report
the discovery of an ultra-relativistic outflow from a binary accreting
neutron star accreting gas within a binary system.  The velocity of
the outflow is comparable to the fastest-moving flows observed from
active galactic nuclei, and its strength is modulated by the rate of
accretion of material onto the neutron star.  Shocks are energized
further downstream in the flow, which are themselves moving with
mildly relativistic bulk velocities and are the sites of the observed
synchrotron emission from the jet. We conclude that the generation of
such highly relativistic outflows does not require properties unique to
black holes, such as an event horizon.}

\newpage

Circinus X-1 is a bright and highly variable X-ray source, containing
a stellar-mass compact object accreting from a binary companion
star. X-ray dips and outbursts with a period of 16.6 days are most
readily interpreted as enhanced accretion during periastron passage of
the accreting object in a significantly elliptical binary orbit
\cite{kal76, jfw99}.  Observations of type-I X-ray bursts indicate
that the accreting object is a neutron star \cite{ten86}. In the 1970s
Cir X-1 was also associated with bright radio outbursts \cite{hay78,
fk01}. Since then, there has been a decline in the strength of the
radio counterpart, which has been found to reside within an extended
radio nebula \cite{hay86}.  Structures on arcminute-scales have been
imaged within this nebula \cite{ste93}, and more recently a one-sided
jet on arcsecond-scales, which aligns with the larger structures, has
been discovered \cite{fen98}.

Bright radio flares associated with the production of a relativistic
jet are now established to be a common property of accreting black
holes in outburst, but are less commonly associated with accreting
neutron stars \cite{fk01}.  In several cases the radio emission has
been resolved into outflowing components, sometimes either side of a
stationary core, with bulk Lorentz factors (the Lorentz factor
$\Gamma = (1-\beta^{2})^{-1/2}$, where velocity $v = \beta c$ and $c$
is the speed of light) typically inferred to be in the range $2 \leq
\Gamma \leq 5$ \cite{mr99}. Based on these observations a generic
qualitative picture has emerged in which electrons are accelerated to
relativistic energies close to the accreting compact object, and then
ejected from the system in a bipolar flow, from which we observe
synchrotron emission until energy losses, primarily due to expansion
of the plasma, reduce the emission below observable levels.  However,
very long baseline interferometric observations of the neutron-star
binary Sco X-1 \cite{fom01a,fom01b} have revealed knots with mildly
relativistic bulk velocities ($\beta\sim0.5$) which are energised by a
beam which is itself not directly detectable, but must have a Lorentz
factor $\Gamma > 3$. Furthermore, it has been suggested that the
observed synchrotron emission results from a moving shock \cite{kai00}
(as discussed for AGN and Gamma-ray bursts), a possibility supported
by recent observations of large-scale X-ray jets from a transient
black hole X-ray binary, revealing {\em in-situ} acceleration of
electrons to TeV energies \cite{cor02}.

\bigskip

At all frequencies and angular scales observed, Cir X-1 is associated
with an extended radio structure.  Careful inspection of Figs 1 and 2
reveals that X-ray flares are associated with a brightening of the
`core' of the extended radio structure. Previous comparison of Hubble
Space Telescope and radio images \cite{fen98} confirms the association
of the radio `core' with the optical counterpart of the binary, and so
we associate the extended emission/knots with an outflow from the
system. The radio spectral indices ($-1.1 \leq \alpha \leq -0.7$)
indicate optically thin synchrotron emission.

We shall focus initially on the observations of October 2000
(left-most panel of Figs 1 and 2), which have a sampling of every
second day. In the `quiescent' observation prior to the X-ray flaring
the radio source at 4.8 GHz is clearly associated with a bright knot
near to the core (best radio position from \cite{fen98}), plus a
weaker extended structure. At 8.6 GHz the extended emission is
resolved into two components, knots A and B, which are not co-aligned
with the core.  Following the X-ray flaring, the core has brightened
significantly within one day, and within three days the extended
emission is brightening, peaking within five days of the
outburst. Subsequently the extended emission fades. In May 2001
(middle panel of Fig 1) we performed a similar set of observations
with two-day sampling and saw a similar effect. Following the X-ray
flaring the extended emission brightens at 4.8 GHz within two or three
days, and there seems to be a `bridge' of radio emission from the core
to the knot immediately after the flare; in addition in the difference
maps (Fig 2) there is some indication of a counterjet.  In Dec 2002 we
made a sequence of observations with {\em daily} sampling. This time
following the X-ray flare the core brightened strongly and a resolved
bridge of radio emission appeared along the apparently curved axis of
the extended emission.

At all epochs we clearly see evidence for the transmission of energy
from the binary core to the extended radio emission following an
outburst. To what apparent velocity does this signal correspond ? 
For a proper motion $\mu$, the apparent velocity expressed as a
fraction of the speed of light $\beta_{\rm app} = v_{\rm app}/c$, for
the source at a distance $d$ is

\[
\beta_{\rm app} \sim \left(\frac{d}{\rm kpc}\right) \left(\frac{\mu}{\rm 170 mas/d}\right) 
\]

All the maps show the brightening of the extended components within
days of the X-ray flare. At all epochs the extended emission at 4.8
GHz, at a distance of 2.2--2.5 arcsec from the core, brightens within
three to five days.  From these numbers we can estimate a {\em
conservative lower limit} to the proper motion associated with the
energising beam of 400 mas d$^{-1}$.  The most likely distance to Cir
X-1 is $\sim 6.5$ kpc \cite{ste93}. From this we can establish that
the apparent signal propagation velocity between the core and knot is
$\beta_{\rm app} > 15$.  For an intrinsic velocity $\beta$ at an angle
$\theta$ to the line of sight,

\[
\beta_{\rm app} = \frac{\beta \sin \theta}{(1-\beta \cos \theta)} 
\]

which has a maximum value of $\beta_{\rm app} = \Gamma \beta$ when
$\beta = \cos \theta$. Since the observations indicate $\beta_{\rm
app} > 15$, this implies $\Gamma \beta > 15$, for which the minimum
velocity solution is $\Gamma > 15$ (i.e. $\beta > 0.998$). This is by
far the most relativistic flow observed to date within our Galaxy.
Furthermore, for {\em any} value of $\beta$, solutions may only be
obtained for angles to the line of sight $\theta < 5^{\circ}$,
indicating that the signal responsible for the brightening of the
knots must propagate at an angle very close to the line of
sight. Furthermore, the images, in particular at 8.6 GHz, give a clear
indication of bending of the jet; by comparison with AGN we can
also infer a small angle to the line of sight, in which case
projection effects increase the apparent bend angle.

This lower limit tothe Lorentz factor may correspond to the bulk
velocity of the flow between the core and the jets, or of a pattern
(shock) propagating along this flow. In the latter case, in the
approximation of a 1D relativistic shock propagating along a
relativistic fluid, the bulk ($\Gamma_{\rm bulk}$) and shock
($\Gamma_{\rm shock}$) Lorentz factors are related by $\Gamma_{\rm
bulk} = \Gamma_{\rm shock} / \sqrt{2}$ in the limit (appropriate here)
of $\beta_{\rm shock} \rightarrow 1$ (e.g. \cite{kro99}). Thus our
lower limit to the signal Lorentz factor of $\Gamma \geq 15$ may place
a lower limit on the {\em bulk} Lorentz factor of the flow of
$\Gamma_{\rm bulk} \geq 10$.

Further clues to the nature of the sites of emission are provided by
linear polarisation maps, to be presented elsewhere, which reveal a
magnetic field in the knots which is parallel to the jet axis in the
core and perpendicular to the jet axis at the knots.  This supports an
interpretation of the extended emission as synchrotron emission from
shocks formed at a longitudinal compression of the jet flow, as in AGN
\cite{ss88}.  Furthermore, it is clear that there is secular evolution
of the mean radio structure between the epochs, with a general trend
for the propagation of components away from the binary core.

\bigskip

These observations reveal that underlying the observed radio jets of
the accreting neutron star Cir X-1 is a flow with a bulk Lorentz
factor $\geq 10$. In another neutron star binary, Sco X-1 the
underlying physics may be similar \cite{fom01a,fom01b}. Sco X-1 is the
prototype of the class of `Z sources', representing the six most
luminous neutron star accretors in our Galaxy, and it has further been
suggested that Cir X-1 itself shares some of the properties of this
class \cite{shi99}. Thus it seems plausible that production of such
ultrarelativistic jets may be a generic feature of accretion onto
neutron stars at or near to the Eddington limit -- this is at
odds with the popular suggestion that the velocities of jets from
neutron stars (and indeed all accreting objects) would be limited to
the surface escape speed, in this case $\sim 0.3c$ \cite{liv99}.

The minimum energy associated with brightening of the extended
emission is $E_{\rm min} \sim 10^{40}$ erg for a source volume of
radius one light day (the estimated energisation timescale); the
associated power input rate is $P_{\rm J} \geq 10^{35}$ erg
s$^{-1}$. If the unseen energising signal were isotropic -- for
example a spherical blast wave from the binary -- this would
correspond to a total power output rate of $\geq 7 \times 10^{39}$ erg
s$^{-1}$, more than an order of magnitude above the Eddington limit
for a neutron star. Based on these arguments we interpret the
signal as arising in a collimated ultrarelativistic flow. 

The velocities of jets from black holes in our galaxy are not well
determined; in particular it is almost impossible to place an upper
limit on the Lorentz factors of the observed motions \cite{fen03}.  In
the case of AGN it has been argued that the bulk Lorentz factor of
motion is $\Gamma \sim 10$ \cite{ghi93}, although in a small fraction
of blazars apparent velocities $\beta \geq 30 h^{-1}$ (where the
Hubble constant $H_0 = 100 h$ km s$^{-1}$ Mpc $^{-1}$) have been
measured on VLBI scales\cite{jor01}, with the same implications for
the underlying bulk velocity as derived here for Cir X-1. Thus these
observations demonstrate that the study of jets from galactic XRBs,
whether containing neutron star or black hole accretors, may be
directly applied to our study of jets from the most powerful engines
in the universe, the AGN. Furthermore, the discovery of such
ultrarelativistic flows associated with a neutron star points us
clearly to the common features of neutron star and black hole
accretion, e.g. the accretion flow, and not properties unique to black
holes, e.g. an event horizon or ergosphere, as the sites of the key
physics in jet formation.

{}                          

\newpage
\newpage

\setcounter{figure}{0}
\begin{figure}
\centerline{\epsfig{file=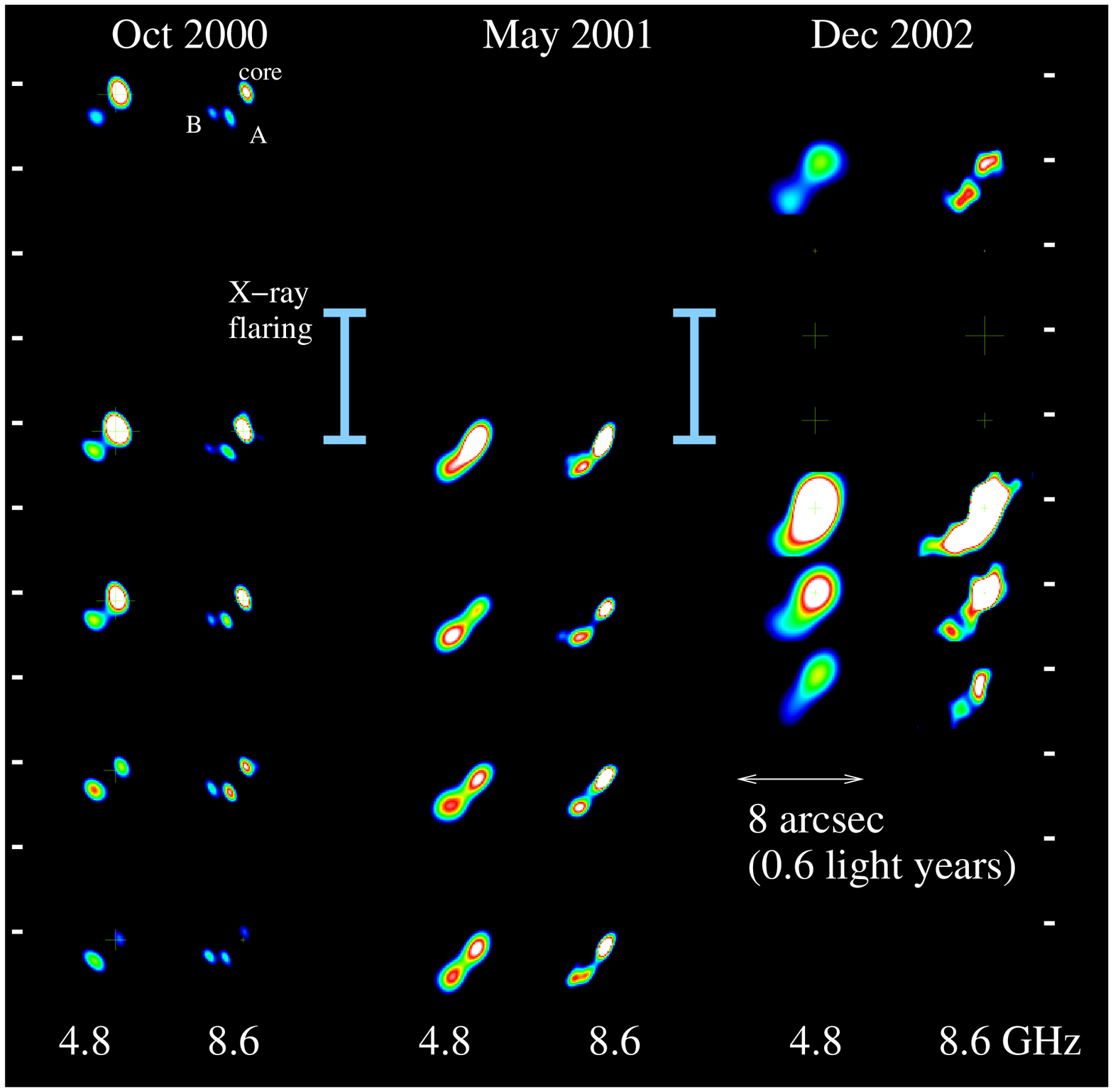, width=14cm, angle=270, clip}}
{\small 
{\bf Fig 1} 
An ultrarelativistic outflow: sequences of radio observations
of Circinus X-1 in October 2000, May 2001 and December 2002. At each
epoch, observations were made simultaneously at 4.8 and 8.6 GHz. White
tickmarks indicate time steps of one day; the blue bar indicates the
time of the X-ray flaring as observed by the {\em Rossi} X-ray Timing
Explorer All-Sky Monitor. In October 2000 and May 2001 the
observations were spaced every two days; in December 2002 they are
daily. At each epoch the {\em u-v} coverage of the radio observations
is identical for each image; maps in October 2000 and May 2001 are
`uniformly weighted', those in December 2002 are `naturally
weighted'. The crosses indicate the location of the binary `core' from
\cite{fen98}, and their size is proportional to the `core' radio flux
density.  The observations reveal that following the X-ray flaring the
extended radio structure brightens on timescales of days. The apparent
velocities associated with this expansion are $\geq 15c$, indicating
an underlying ultrarelativistic flow.  }
\end{figure}

\begin{figure}
\centerline{\epsfig{file=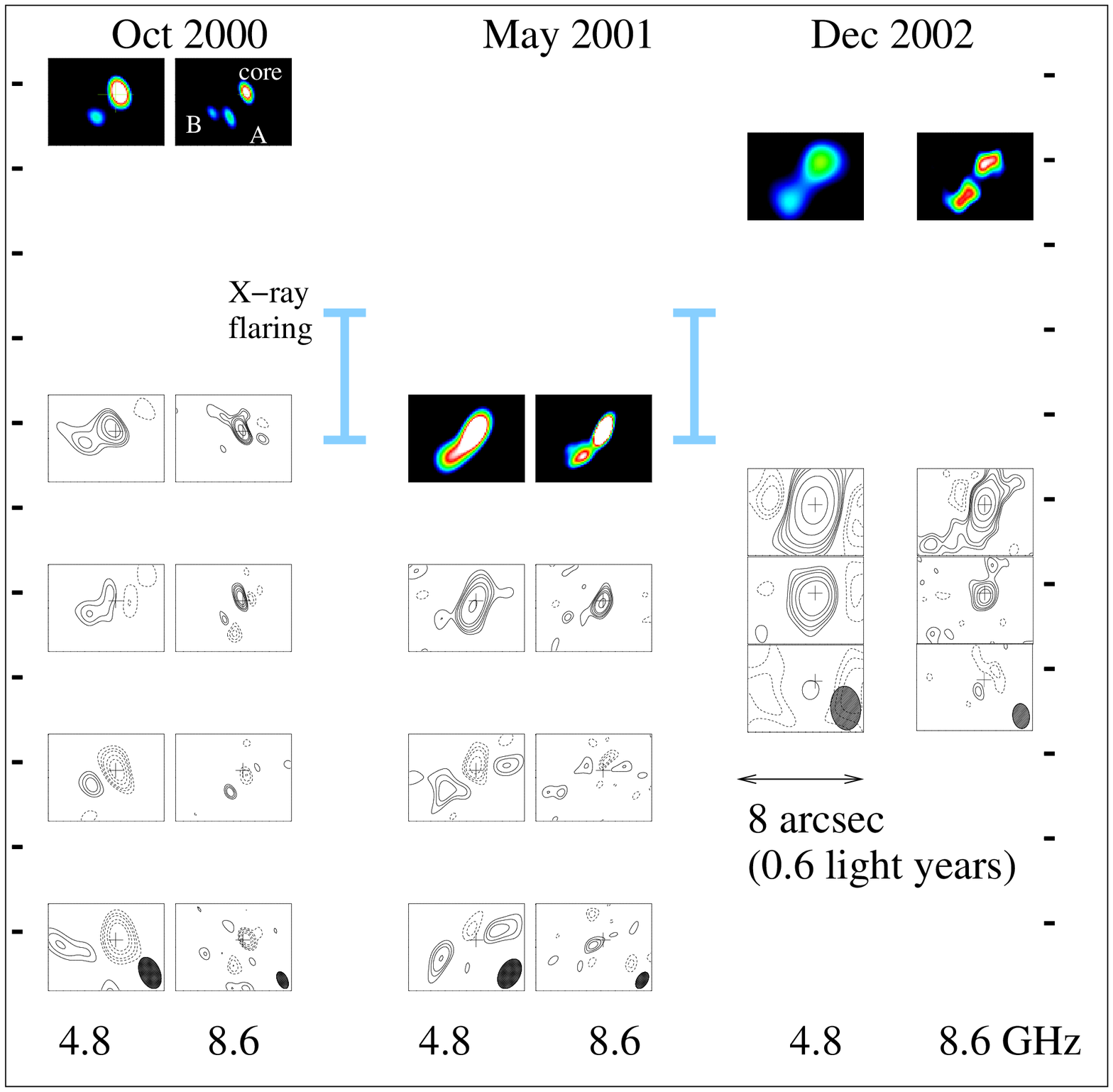, width=14cm, angle=270, clip}}
{\small 
{\bf Fig 2} 
The same data as Fig 1, but presented in the form of
`difference' maps. At each of the three epochs the top (reference)
image at each frequency is the same as that presented in Fig 1. The
contour images below it correspond to the maps from the corresponding
frames in Fig 1 with the reference image subtracted, in order to
clearly illustrate changes in the radio structure. Solid lines
indicate positive residuals, dashed lines indicate negative residuals;
contour levels have been chosen to indicate the level of noise in each
image. At each epoch, the strongest variability in the difference maps
is clearly occurring in the region of the jet / knots. Solid ellipses
in the lower right corners of the bottom panels indicate the size and
shape of the synthesised beam at each epoch and frequency.}
\end{figure}

\end{document}